\documentclass[a4paper]{article}

\usepackage{booktabs}
\usepackage{graphicx}
\usepackage[utf8]{inputenc}
\usepackage{amsmath}
\usepackage{amssymb}
\usepackage{cancel}
\usepackage{xcolor}
\usepackage{tikz}
\usepackage{url}
\usepackage{subcaption}
\usepackage{buildinggraph}
\usepackage[ruled]{algorithm2e}
\usepackage[numbers]{natbib}

\usetikzlibrary{positioning,shapes,calc,through,intersections,fit,matrix,patterns}

\tikzset{
  flowchart/.style={
      process/.style={rounded rectangle, draw,align=center},
      decision/.style={diamond, aspect=2, draw,align=center},
      transition/.style={draw,->},
      dataflow/.style={->,dashed},
    },
    mnemonic/.style={
      label={[fill=blue!30]120:#1}
    },
    land/.style={fill=green!30!yellow!30, draw=none},
    sea/.style={fill=blue!60!green!20, draw=none},
}

\allowdisplaybreaks[1]

\title{Anchoring Theory in Sequential Stackelberg Games}
\author{Jan Karwowski \and Jacek Ma{\'n}dziuk \and Adam \.Zychowski}

\begin{document}

\maketitle
\begin{abstract}
An underlying assumption of Stackelberg Games (SGs) is perfect rationality of the players. However, in real-life situations (which are often modeled by SGs) the followers (terrorists, thieves, poachers or smugglers) -- as humans in general -- may act not in a perfectly rational way, as their decisions may be affected by biases of various kinds which bound rationality of their decisions. One of the popular models of bounded rationality (BR) is Anchoring Theory (AT) which claims that humans have a tendency to flatten probabilities of available options, i.e. they perceive a distribution of these probabilities as being closer to the uniform distribution than it really is. This paper proposes an efficient formulation of AT in sequential extensive-form SGs (named \emph{ATSG}), suitable for Mixed-Integer Linear Program (MILP) solution methods. ATSG is implemented in three MILP/LP-based state-of-the-art methods for solving sequential SGs and two recently introduced non-MILP approaches: one relying on Monte Carlo sampling (\emph{O2UCT}) and the other one (\emph{EASG}) employing Evolutionary Algorithms.
Experimental evaluation indicates that both non-MILP heuristic approaches scale better in time than MILP solutions while providing optimal or close-to-optimal solutions.
Except for competitive time scalability, an additional asset of non-MILP methods is flexibility of potential BR formulations they are able to incorporate. While MILP approaches accept BR formulations with linear constraints only, no restrictions on the BR form are imposed in either of the two non-MILP methods.
\end{abstract}

\section{Introduction}%
\label{sec:introduction}

Stackelberg Games (SGs)~\cite{Leitmann_1978,stackelbergmarktform} are a game-theoretic model which attracted considerable interest in recent years, in particular in Security
Games area~\cite{sinha2018securitygames}. In its simplest form Stackelberg Security Game (SSG) assumes two players: a \emph{leader} who commits to a (mixed) strategy first,
and a \emph{follower} who makes their commitment already knowing decision of the leader.
The above asymmetry of the players very well corresponds to interactions between law enforcement forces (leaders) and smugglers, terrorists or poachers (followers) modeled by SSGs~\cite{fang2015security,jain2010software,yin2012,shieh2012protect}.

A fundamental assumption in SGs is that the follower will make an optimal, perfectly rational decision exploiting knowledge about the leader's commitment. However, in real-life
scenarios the follower may suffer from cognitive biases or bounded rationality leading them to suboptimal decisions~\cite{yang2013improving,cobra,fang2015security}.

On a general note, \emph{bounded rationality} (BR)~\cite{simon1957models} in problem-solving refers to limitations of decision-makers that lead them to taking non-optimal actions. Except for limited cognitive abilities, BR can be attributed to partial knowledge about the problem, limited resources, or imprecisely defined goal~\cite{aumann1997rationality,rubinstein1998modeling}.
The most popular models of BR are Prospect Theory (PT)~\cite{kahneman2013prospect}, Anchoring Theory (AT)~\cite{tversky1974judgment}, Quantal Response (QR)~\cite{mckelvey1995quantal} and Framing Effect (FE)~\cite{tversky1981framing}. Each of these models has specific problem-related assumptions and each of them possesses certain experimental justification, though none of them could be regarded as a widely-agreeable leading BR formulation.

The concept of BR plays an important role in SSGs, as in their real-world applications the follower's role is played by humans, e.g.\ terrorists, poachers or criminals, who usually suffer from BR limitations.
One of the most popular BR approaches in SSG domain is COBRA~\cite{cobra,pita2010robust} which modifies DOBSS MILP (Mixed-Integer Linear Program)~\cite{paruchuri2008playing} to address AT with $\epsilon$-optimality models.
A similar approach was taken by Yang et al.~\cite{yang2013improving} who proposed BR models relying on PT and QR, resp. and demonstrated their suitability in SSGs based on experiments involving humans.
SHARP system~\cite{kar2015game} points certain game-related aspects (e.g. past performance, similarity between game conditions, etc.) which need to be taken into account by an algorithm for repeated SSGs when playing against human adversaries.
MATCH approach~\cite{pita2012robust} optimizes the leader's strategy against a worst-case outcome within some error bound (i.e. assuming certain deviations from the follower's optimal strategy).
Another approach - BRQR~\cite{yang2011improving} proposed by Yang et al. is based on the idea of QR, further improved in SU-BRQR system~\cite{nguyen2013analyzing} which introduces a subjective utility function for the follower with parameters tuned in the experiments involving humans.
Despite clear variability, all existing implementations of BR in SSG domain employ MILP for finding the game solution (Stackelberg Equilibrium) and all of them are limited to one-step (non-sequential) games.

In this paper, AT approach implemented in COBRA~\cite{cobra,pita2010robust} for single-step normal-form games is extended to the case of \emph{sequential extensive-form games}
in a way that avoids non-linear constraints, which makes it suitable for a wide range of MILP/LP approaches. Consequently, modifications to three state-of-the-art methods for solving extensive-form SSGs~\cite{cermak2016using,cerny2018incremental,bosansky2015} which implement AT are proposed. Furthermore, two other non-MILP heuristic methods for solving SSG that rely on Monte Carlo sampling~\cite{KarwowskiMandziuk2019,O2UCT} and Evolutionary Algorithm~\cite{EASG}, respectively are also adequately modified to incorporate AT principles. All five methods are experimentally evaluated on a set of Warehouse Games~\cite{EJOR2019,sg-mini}.

\subsection{Definitions}%
\label{sec:definitions}
Throughout the paper a notation from~\cite{cermak2016using} will be used so as to easily refer to the method proposed in that paper. Sequential games will be represented as \emph{Extensive-Form Games~(EFGs)}, i.e. tuples $G=(\mathcal{N}, \mathcal{H}, \mathcal{Z}, \mathcal{A}, \rho, u, \mathcal{I})$, where $\mathcal{N}=\{l,f\}$ is a set of players, the leader and the follower respectively. $\mathcal{H}$ is a set of game nodes which compose a game tree with a root node representing the initial game position. $\mathcal{Z}\subset \mathcal{H}$ is a set of leaves representing terminal game states. $\mathcal{A}$ is a family of sets $A_h$, which $\forall h\in\mathcal{H}\setminus\mathcal{Z}$ define possible actions from each non-terminal node. $\rho: \mathcal{H}\setminus\mathcal{Z} \rightarrow \mathcal{N}$ is a function that defines an acting player in a given node. $u = \{u_l, u_f\},\ u_i : \mathcal{Z} \rightarrow \mathbb{R}, i\in \mathcal{N}$ is a family of utility functions that assign a game outcome in a terminal node to the respective player. $\mathcal{I}$ is a family of Information Sets~(ISs); each $I\in \mathcal{I}$ defines states that are indistinguishable to the acting player. $\mathcal{I}$ satisfies the following conditions:
  \begin{itemize}
  \item $\mathcal{I}$ partitions $\mathcal{H}\setminus \mathcal{Z}$,
  \item $\forall I\in\mathcal{I}\quad \forall h_1, h_2 \in I \quad \rho(h_1)=\rho(h_2)$ -- all nodes in a given IS have the same acting player,
  \item $\forall I\in\mathcal{I}\quad \forall h_1,h_2 \in I \quad A_{h_1}=A_{h_2}$ -- the set of possible actions is the same for all nodes from a given IS.
\end{itemize}
Additionally, $A(I)$ will denote the set of actions available in $I$ and $\mathcal{I}_i, i\in\mathcal{N}$ a family of ISs with acting player $i$ ($\mathcal{I}=\mathcal{I}_l \cup \mathcal{I}_f$).

Moreover, the games are assumed to satisfy the \emph{perfect recall} property,
i.e. throughout the game each player is fully aware of previous ISs visited by them and actions taken in that ISs.

In EFG a \emph{pure strategy} of a player assigns to each IS in which
the player is an acting one a particular action to be played in that IS.
A \emph{mixed strategy} of a player is a probability distribution over
pure strategies of that player. $\Pi_i$ / $\Delta_i$ will denote a set
of pure / mixed strategies of player $i$, resp. Elements of $\Pi_i$
and $\Delta_i$ will be denoted by $\pi$ and $\delta$ with
adequate indices, resp.

A \emph{behavior strategy} is an assignment of a probability distribution of actions for each
IS that a player can reach during the game. It can be viewed as a tree with nodes representing player's ISs and edges representing actions (labeled
with their probabilities).
The notions of \emph{mixed strategy} and \emph{behavior strategy} will be used interchangeably as they are equivalent in games with perfect recall.

We will overload the notation of $u$ functions, so that $u_i(\pi_l, \pi_f)$ would denote the $i$-th player's utility after the pure strategy profile $(\pi_l, \pi_f)$ was played. Similarly,
$u_i(\delta_l, \delta_f)$ will denote the expected utility value of the $i$-th player in reference to the mixed strategy profile $(\delta_l, \delta_f)$.

Each node in a game tree is uniquely defined by a pair of sequences: the leader's actions and the follower's actions which lead to that node. These sequences will be denoted by $\sigma_l$ and $\sigma_f$, resp. We will say that a pair of sequences $(\sigma_l, \sigma_f)$ is \emph{compatible} if it leads to a terminal node in a game tree.
Utility values in terminal nodes pointed by a compatible pair of sequences will be denoted by $u_i(\sigma_l, \sigma_f), i\in\mathcal{N}$.
Following~\cite{cermak2016using}, for any pair $(\sigma_l, \sigma_f)$ we will define an auxiliary function $g_i(\sigma_l, \sigma_f)$ which yields a value of $u_i(\sigma_l, \sigma_f)$ if the sequences are compatible and $0$ otherwise.

Finally, $\sigma_i(h), i\in \mathcal{N}, h\in \mathcal{H}$ will denote a sequence of moves of the $i$-th player which led to node $h$ and $I_i(\sigma_i)$ the IS in which the last action from $\sigma_i$ was played.

The goal of SG is to find \emph{Stackelberg Equilibrium (SE)}, i.e. a strategy profile $(\delta_l, \delta_f)$ that is a solution of the following set of equations:
\begin{equation}
 \begin{cases}
    \max_{\delta_l} u_l(\delta_l, BR(\delta_l))\\
    BR(\delta_l) = \text{argmax}_{\delta_f} u_f(\delta_l, \delta_f)
  \end{cases}
\label{eq:SE}
\end{equation}

Please observe that SE is not well defined when there is more than one best follower's response. For this reason SE is often extended to the form of \emph{Strong Stackelberg Equilibirum~(SSE)}~\cite{Breton_1988} in which, in addition to (\ref{eq:SE}), in the case of a tie among follower's best response strategies, the of them that maximizes the leader's utility is selected (if there are more such strategies anyone of them is chosen). The SSE version of SE is considered in this paper.

\subsection{Motivation}%
\label{sec:anchoring-modification}
The paper combines the following two concepts, which are generally considered separately in the literature: (1) bounded rationality models in Security Games and (2) efficient solutions for sequential SGs. In both areas significant progress has been observed in recent years.

To our knowledge, the concept of BR has been addressed in Security Games only in the context of single-step games, for instance, in the recently emerged, fast-growing genre of Green Security Games~\cite{xu-blackbox,fang2015security}, in which game theoretical models exploit not rational behavior of attackers (e.g. poachers or illegal forest extractors) to maximize the effectiveness of protection activities.

At the same time, in reference to large-scale sequential SGs, several algorithms utilizing different techniques, e.g. sequence-form~\cite{bosansky2015}, correlated equilibria~\cite{cermak2016using}, game abstraction~\cite{cerny2018incremental}, Evolutionary Algorithm~\cite{nasza_AAAI,EASG} or Monte Carlo sampling~\cite{EJOR2019,O2UCT} which visibly extended the range of tractable SGs, have been proposed recently.

We believe that successful studies on the crossroads of these two research directions will allow for tackling large-scale problems that better (more realistically) model security situations involving humans.

Among several BR models introduced in the literature, we have chosen the AT approach since it was already successfully applied to single-step SGs~\cite{cobra} and, furthermore,
is intuitively justified in the cases when only limited observation of the leader's strategy is possible.

Generally speaking, AT~\cite{tversky1974judgment} assumes the existence of a bias of a person who observes some events (for instance, surveils the opponent's strategy in SSG) towards the uniform distribution. Formally, for any probability distribution over a finite set $X$, let us denote the probability of $x\in X$ as $q_x$. The observer believes that this probability is equal to $q_x'=q_x(1-\alpha)+\alpha/|X|$, where $0<\alpha <1$ is a parameter of AT bias and $|X|$ is cardinality of $X$.

In SGs the leader, being aware of the follower's AT bias, can exploit this knowledge in their mixed strategy formulation.

\subsection{Contribution}%
\label{sec:contribution}
The main contribution of this paper can be summarized as follows:
\begin{itemize}
\item Introduction of efficient MILP-suitable extension of Anchoring Theory to the genre of sequential (multi-step) Stackelberg Games (ATSG);
\item Implementation of ATSG in three MILP/LP-based state-of-the-art methods for solving sequential Stackelberg Games (two exact and one approximate) and in two approximate non-MILP approaches, relying on Monte Carlo sampling and Evolutionary Algorithm, respectively;
\item Experimental evaluation of five above-mentioned methods
in BR settings with respect to the quality of payoffs and time efficiency.
\end{itemize}

\section{Anchoring Theory in Sequential Games}%
\label{sec:at-seq}

As we mentioned above, implementations of AT in SGs presented in the literature are limited to single-step games only.
There are two straightforward ways to generalize AT to sequential games.

The first one is to transform an extensive-form game to its normal-form where each player's actions are equivalent to pure strategies. Such an approach, however, would introduce a \emph{global distortion} of probabilities and is, therefore, inaccurate when the opponent's behavior is considered separately in each decision point.

The other possibility is to apply AT distortion \emph{locally} - i.e. to a probability distribution in each IS that forms player's behavior strategy. Such an approach seems to be more intuitive, especially considering the fact that a behavior strategy is usually a natural way of perceiving a mixed strategy by humans. Unfortunately, due to non-linear constraints, such a generalization poses problems for sequence-form MILP methods.

The following subsection presents a solution that yields similar distortion to the one described above while avoids non-linear constraints.

\subsection{Sequence-Form based MILPs}
\label{sec:C2016}
A state-of-the-art approach to calculate SSE in sequential games~\cite{cermak2016using} -- referred to as \emph{C2016} in this paper -- is an iterative method which alternates two phases: solving MILP/LP for finding Stackelberg Extensive-Form Correlated Equilibrium (SEFCE) in sequence-form game representation, and SEFCE refinement with a dedicated procedure relying on LP modification towards SSE. Since the equilibrium refinement part is not affected by the implementation of AT it will not be discussed here. The SEFCE part of the method, defined by a set of equations (\ref{eq:milp-start})-(\ref{eq:milp-end}), is built around variables that define probabilities of playing particular sequences by the players~\cite{cermak2016using}.
The following LP definition employs the notion of relevant sequence pairs ($rel$) -- formally introduced in Definition~3 of~\cite{cermak2016using}.
\begin{align}
\max_{p,v} & \sum_{\sigma_l\in\Sigma_l}\sum_{\sigma_f\in\Sigma_f}
  p(\sigma_l, \sigma_f)g_l(\sigma_l,\sigma_f)   \label{eq:milp-start}\\
\text{s.t.}\qquad & p(\emptyset, \emptyset) = 1; 0 \leq p(\sigma_l,
              \sigma_f) \leq 1 \label{eq:plan1} \\
p(\sigma_l(I), \sigma_f)=&\hspace{-.3cm}\sum_{a\in A(I)}\hspace{-.2cm}p(\sigma_l(I)a, \sigma_f)
                           \;\; \forall I\in \mathcal{I}_l, \forall
                           \sigma_f \!\in\! rel(\sigma_l(I))  \label{eq:plan2}\\
p(\sigma_l, \sigma_f(I)) =&\hspace{-.3cm}\sum_{a\in A(I)}\hspace{-.2cm} p(\sigma_l,
                            \!\sigma_f(I)a)\;\; \forall I \!\in\!
                            \mathcal{I}_f,\forall \sigma_l \!\in\!
                            rel(\sigma_f(I))  \label{eq:plan3}\\
v(\sigma_f)=& \sum_{\sigma_l \in rel(\sigma_f)} p(\sigma_l,
  \sigma_f)g_f(\sigma_l, \sigma_f) \notag \\+& \sum_{I\in
              \mathcal{I}|\sigma_f(I)=\sigma_f} \sum_{a\in A_f(I)} \hspace{-.4cm}
              v(\sigma_f a)\quad \forall \sigma_f\in\Sigma_f \label{eq:foll-constr1}\\
v(I,\sigma_f) \geq& \sum_{\sigma_l \in rel(\sigma_f)} p(\sigma_l,
  \sigma_f)g_f(\sigma_l, \sigma_f(I)a) \notag \\+& \sum_{I'\in \mathcal{I}_f |
                    \sigma_f(I')=\sigma_f(I)a} v(I', \sigma_f), \notag
  \\ &
                    \hspace{-.6cm}\forall I \in \mathcal{I}_f, \forall{\sigma_f}
                    \in \bigcup_{h\in I} rel(\sigma_l(h)), \forall a
                    \in A(I)  \label{eq:foll-constr2} \\
v(\sigma_f(I)a)=&v(I, \sigma_f(I)a) \forall I \in \mathcal{I}_f,
  \forall a\in A(I) \label{eq:milp-end}
\end{align}
The main variables in the above LP are $p(\sigma_l, \sigma_f)$ which describe the correlation plan and represent probabilities that correlation device will give suggestion of playing the respective sequences of moves $(\sigma_l, \sigma_f)$ by the players. Implicitly, they define the resulting players' strategies. Objective~(\ref{eq:milp-start}) maximizes leader's utility. Constraints~(\ref{eq:plan1}) -- (\ref{eq:plan3}) ensure that the correlation plan is correct, i.e.\ probability of playing a given sequence is a sum of probabilities of playing sequences that are built from it by adding one action. $v$ are auxiliary variables which guarantee that suggested $\sigma_f$ is the best follower's response.
The crucial constraints (from AT perspective) are~(\ref{eq:foll-constr1}) and~(\ref{eq:foll-constr2}) which assure that the selected follower's strategy would yield an outcome no worse than that of any other strategy. Implementation of AT requires changing the perception of $p(\sigma_l, \sigma_f)$ variables so as to include anchoring bias - the details are presented in the following section.

Solving the above LP is iteratively alternated with a refinement procedure mentioned above. No modifications to this procedure, compared to its original formulation~\cite{cermak2016using}, are required.

\subsection{Anchoring Theory modification}%
\label{sec:at-modification}

ATSG is implemented as a distorted follower's perception of the leader's behavior strategy. Let's denote by $q(i)$ a probability of choosing action $i$  by the leader in a given IS, stemming from its behavior strategy.
The most straightforward implementation of AT (though non-linear in sequence-form games) is to change the probability of taking this action to $q'(i) = (1-\alpha q(i)) + \alpha /M$, where $M$ is the number of actions available in this IS. However, in sequence-form games, for a given leader's sequence of actions $\sigma_l=a_1,a_2,a_3,\ldots,a_n$ a probability of playing it, based on behavior strategy, would be $p(\sigma_l)=q(a_1)q(a_2)\cdots q(a_n)$ and the distorted AT probability would become
\begin{align}
p'(\sigma_l)= (&(1-\alpha) q(a_1) + \alpha /M_1) ((1-\alpha) q(a_2)
                  + \alpha /M_2) \cdots \notag \\(&(1-\alpha) q(a_n) + \alpha /M_n),\label{eq:ATSG-nl}
\end{align}
\noindent
where $M_i$ is the number of actions available in IS in which $a_i$ is played.

Please observe that variables $p$ in LP formulation (\ref{eq:milp-start})-(\ref{eq:milp-end}) are products of $q(a_i)$ values presented above (\ref{eq:ATSG-nl}), and as such cannot be expressed in a linear form with respect to $q(a_i)$. Consequently, applying the above AT modification to MILP~(\ref{eq:milp-start})--(\ref{eq:milp-end}) would end-up with non-linear constraints, inadequate for MILP formulation.

Consequently, we propose to simplify the above ATSG by dropping distortion coefficients from all but the last one probabilities:
  \begin{align}
   &   p''(\sigma_l)= q(a_1)q(a_2)\cdots q(a_{n-1})
  ((1-\alpha) q_{a_n} + \alpha /M_n) \notag\\& =
  q(a_1)q(a_2)\cdots q(a_{n-1})\!\! \cdot\! \alpha
  /M_n+(1\!-\!\alpha)q(a_1)q(a_2)\!\cdots\!q(a_{n-1})q(a_n) \notag\\& =
  p(init(\sigma_l)) \alpha/M_n + (1-\alpha) p(\sigma_l),  \label{eq:ATSG}
  \end{align}
\noindent
where $init(\cdot)$ is a function which outputs a sequence without the last move. A simplified version of ATSG (eq.~(\ref{eq:ATSG})) is well suited to MILP/LP formulations of sequence-form games.

Please note that relations among probabilities of the leader's actions
within a single IS are the same according to both
eqs.~(\ref{eq:ATSG-nl}) and~(\ref{eq:ATSG}), i.e. $\forall \sigma_l^1,
\sigma_l^2 \quad I(\sigma_l^1)=I(\sigma_l^2) \Rightarrow
p'(\sigma_l^1)/p'(\sigma_l^2)= p''(\sigma_l^1)/p''_(\sigma_l^2)$, where $p'(\sigma), p''(\sigma)$ denote probability of sequence $\sigma$ in a given IS calculated according to (\ref{eq:ATSG-nl}) and (\ref{eq:ATSG}), resp. Furthermore, for a given sequence $\sigma_l$, for small values of $\alpha$ a difference $|p''(\sigma_l)-p'(\sigma_l)|$ is also small.

Please also note that the resulting $p''$ values do not represent proper probability distribution since they do not sum up to one. Their normalization is not needed though, as they are used only to make comparisons between distorted utilities of various follower's strategies. Results of such comparisons are independent of $p''$ normalization.

\subsection{Modification of MILP/LP based methods}%
\label{sec:mod-lp}

ATSG formulation (\ref{eq:ATSG}) was incorporated into three state-of-the-art methods for sequential SGs.

\subsubsection{SEFCE method} In the first method~\cite{cermak2016using}, briefly summarized in section~\ref{sec:C2016}, ATSG is implemented through modification of constraints~(\ref{eq:foll-constr1}) and~(\ref{eq:foll-constr2}), which are replaced by constraints~(\ref{eq:foll-constr1-mod}) and~(\ref{eq:foll-constr2-mod}) presented below:
\begin{align}
  &v(\sigma_f)=
                \xcancel{\sum_{\sigma_l \in rel(\sigma_f)} p(\sigma_l,\sigma_f)g_f(\sigma_l, \sigma_f)}\notag\\
   &             \sum_{\sigma_l \in rel(\sigma_f)}\hspace{-.3cm}
                g_f(\sigma_l,
                \sigma_f) \left( p(\sigma_l,\sigma_f)+\alpha/M_I\cdot p(init(\sigma_l),
                \sigma_f)\right)  + \notag \\
&              + \sum_{I\in
              \mathcal{I}|\sigma_f(I)=\sigma_f} \sum_{a\in A_f(I)}
              v(\sigma_f a)\quad \forall \sigma_f\in\Sigma_f \label{eq:foll-constr1-mod}\\
&  v(I,\sigma_f) \geq
                      \xcancel{\sum_{\sigma_l \in rel(\sigma_f)} p(\sigma_l,\sigma_f)g_f(\sigma_l, \sigma_f(I)a)}
                      \sum_{\sigma_l \in rel(\sigma_f)} \notag \\ &
                      g_f(\sigma_l, \sigma_f(I)a)\left(p(\sigma_l,\sigma_f)
                      + \alpha/M_I\cdot p(init(\sigma_l), \sigma_f)\right)
                      + \notag\\
                     & + \sum_{I'\in \mathcal{I}_f |
                    \sigma_f(I')=\sigma_f(I)a} v(I', \sigma_f), \notag
  \\ &
                    \forall I \in \mathcal{I}_f, \forall{\sigma_f}
                    \in \bigcup_{h\in I} rel(\sigma_l(h)), \forall a
                    \in A(I)  \label{eq:foll-constr2-mod}
\end{align}

The above formulation is a result of application of eq.~(\ref{eq:ATSG}) to LP constraints.

Please observe that LP in \emph{C2016} does not contain variables describing probabilities of playing $\sigma_l$ alone $(p_{\sigma_l})$, but refers to a correlation plan which provides suggestions on the playing pairs $(\sigma_f, \sigma_l)$. Moreover, $p(\sigma_l, \sigma_f)$ equals $p(\sigma_l)$ only if the correlation plan suggests the follower to play a pure strategy (i.e. marginal probability $p(\sigma_f)\in\{0,1\}$ (*)). In the above ATSG version of \emph{C2016}, defined by equations~(\ref{eq:foll-constr1-mod}) -(\ref{eq:foll-constr2-mod}), conditions (*) may not initially hold for all $\sigma_f$, but must be all fulfilled at completion of \emph{C2016}, since they constitute a stopping condition of this method.

\subsubsection{Game abstraction method} In 2018 a new approach to
extensive-form SSGs that folds game subtrees into nodes called \emph{gadgets}
and then incrementally unfolds them to refine the
solution~\cite{cerny2018incremental} was proposed. The method
(henceforth referred to as \emph{CBK2018}) internally employs
\emph{C2016} to solve the abstracted (smaller) games. \emph{CBK2018} was formulated by its authors in two variants: as an exact method and as a heuristic time-optimized approach, with experimental evaluation provided only for the latter variant~\cite{cerny2018incremental}. Consequently, we also focus on heuristic formulation of \emph{CBK2018} and
following recommendation of the authors of~\cite{cerny2018incremental} set the internal method's parameters to $\epsilon=0.3, \sigma=0.4$ which assures fast convergence, albeit at the cost of some deviation from the optimal results. In ATSG modification of \emph{CBK2018} original \emph{C2016} formulation is replaced with its ATSG version (\ref{eq:foll-constr1-mod})-(\ref{eq:foll-constr2-mod}).

\subsubsection{Sequence-form method}
The third MILP-based method for finding SE in sequential games considered in this paper is an approach proposed in~\cite{bosansky2015} (henceforth referred to as \emph{BC2015}). \emph{BC2015} directly utilizes sequence-form game representation and, unlike \emph{C2016}, is not an iterative method, i.e. relies on solving a single MILP instance to obtain the game solution. Generally, its performance is expected to be worse than \emph{C2016} due to the substantial number of integer variables in MILP (one variable per each possible follower's sequence of moves). Below, a modified MILP which incorporates eq.~(\ref{eq:ATSG}) into \emph{BC2015} formulation is presented:
\begin{align}
      &\max_{p,r,v,s}\sum_{z\in Z}p(z)u_l(z)\text{, s.t.}\\
      v_{I_f(\sigma_f)} =& s_{\sigma_f} + \hspace{-.4cm}
      \sum_{I'\in\mathcal{I}_f|\sigma_f(I')=\sigma_f} \hspace{-.7cm}
                           v_{I'}\quad  + \xcancel{\sum_{\sigma_l \in \Sigma_l} r_l(\sigma_l)g_f(\sigma_l,
      \sigma_f)}+ \notag \\
      & \hspace{-1.5cm} \sum_{\sigma_l \in \Sigma_l} \hspace{-.1cm}  r_l(\sigma_l)g_f(\sigma_l,
      \sigma_f)\! + \!\alpha/M_{I(\sigma_l)} g_f(\sigma_l,
      \sigma_f) r_l(init(\sigma_l)) \\
      r_i(\emptyset) =& 1 \quad \forall i\in N \\
      r_i(\sigma_i) =& \!\!\!\!\!\! \sum_{a\in A_i(I_i)}\!\!\!\! r_i(\sigma_i a) \quad
                       \forall i \in N \;\forall I_i \in \mathcal{I}_i, \sigma_i = \sigma_i(I_i) \\
      0 \leq& s_{\sigma_f} \leq (1 - r_f(\sigma_f)) \cdot M \quad \forall \sigma_f \in \Sigma_f\\
      0 \leq& p(z) \leq r_i(\sigma_i(z)) \quad \forall i \in N \quad \forall z\in Z \\
      1 =& \sum_{z\in Z}p(z)  \\
      &r_f(\sigma_f) \in \{0, 1\} \quad \forall \sigma_f \in \Sigma_f \notag \\
      &0 \leq r_l(\sigma_l) \leq 1 \quad \forall \sigma_l \in \Sigma_l \notag
\end{align}

\section{Heuristic Approximations of ATSG }%
\label{sec:approximation}
The above three ATSG modifications of MILP/LP methods are compared with two heuristic non-MILP approaches to solving sequential extensive-form SSG with adequate ATSG adjustments.
\subsection{A summary of \emph{O2UCT} method}%
\label{sec:o2uct}

The first approach (referred to as \emph{O2UCT} - double-oracle UCT sampling)~\cite{KarwowskiMandziuk2019,O2UCT} relies on a guided sampling of the follower's strategy space interleaved with finding a feasible leader's strategy using double-oracle method.

In the first step, a follower's strategy ($\pi_{f}^{r}$) is obtained using Upper Confidence bound applied to Trees (UCT) algorithm~\cite{UCT} - a variant of guided Monte Carlo sampling. Then, for the sampled follower's strategy, a process of building the leader's strategy ($\delta_l$) is performed. $\delta_{l}$
must satisfy the following conditions: (1) $\pi_{f}^{r}$ is the best response strategy against $\delta_{l}$; (2) $\delta_{l}$ provides as high as possible leader's utility when played against the best follower's response. An algorithm of finding the requested leader's strategy $\delta_{l}$ is
outlined below and detailed in~\cite{O2UCT}.

In the first step the best follower's response ($\pi_{f}^{b}$) is calculated ($\dagger$) against $\delta_l$. Then the algorithm checks if the $\pi_{f}^{b}=\pi_{f}^{r}$. If so, then the procedure for adjusting $\delta_l$ to obtain better utility against $\pi_{f}^{b}$ (compared to $\pi_{f}^{r}$) is applied ($\ddagger$). Otherwise, when $\pi_{f}^{r}\neq\pi_{f}^{b}$, an adjustment to $\delta_l$ is made so as to increase leader's utility against $\pi_{f}^{r}$.

The two above-mentioned phases: sampling of the follower's strategy $\pi_{f}^{r}$ (against the current leader's strategy $\delta_l$) and adjustment of $\delta_l$ are iteratively alternated in \emph{O2UCT}.

\subsubsection{ATSG implementation} ATSG implementation in
\emph{O2UCT} required two changes. In the follower's best response
oracle ($\dagger$), which works by exhaustive search of possible pure
strategies in \emph{O2UCT}, the procedure that calculates follower's
utility was modified so as to use distorted probabilities (\ref{eq:ATSG})
when calculating the expected value. Similarly, in the procedure that
calculates a difference between follower's utilities for two
strategies ($\ddagger$), the way the expected utility is calculated was adapted so as to use a distorted strategy (perceived by the follower in ATSG).

Please observe that in the case of \emph{O2UCT}, contrary to MILP/LP ATSG implementations, the potential existence of non-linearities in the formulas defining distorted follower's probabilities is not harmful, and - in principle - any other BR modification could be used instead of eq.~(\ref{eq:ATSG}). For comparability reasons, we will use a linear form (\ref{eq:ATSG}) in the experiments.

\subsection{A summary of \emph{EASG} method}%
\label{sec:ea}
The other heuristic method applicable to sequential SGs considered in this paper~\cite{EASG} utilizes EA to find the leader's mixed strategy and, to our knowledge, is the first generic evolutionary approach proposed in this domain. We are aware of only one other application of EAs to solving sequential SGs~\cite{nasza_AAAI} which, however, is specifically designed to games on a plane with moving targets.

\begin{algorithm}
\small
  \SetNoFillComment
  $\mathcal{P}$ - population\\
  $\mathcal{P} \gets$ randomly selected leader's pure strategies\\
  \While{(generations limit is not reached)}{
  $E \gets$ $n_e$ chromosomes with the highest fitness function values\\
  $\mathcal{P}_c \subseteq \mathcal{P}$ \tcc{random population subset for crossover}
  \tcc{crossover merges pairs of chromosomes}
  $\mathcal{P} = \mathcal{P} \cup Crossover(\mathcal{P}_c)$\\
  $\mathcal{P}_m \subseteq \mathcal{P}$ \tcc{random population subset for mutation}
  \tcc{mutation changes actions in randomly selected element of a chromosome}
  $\mathcal{P} = (\mathcal{P}\setminus\mathcal{P}_m) \cup Mutation(\mathcal{P}_m)$\\
  $Evaluate(\mathcal{P})$ \tcc{calculate fitness function value - the leader's payoff against the optimal follower's response to a strategy encoded in a chromosome}
  $\mathcal{P} = E \cup Selection(\mathcal{P})$ \tcc{choose strategies for the next generation based on fitness evaluation}
  }
  \Return{best leader's strategy}
    \caption{A pseudocode of \emph{EASG}.}
    \label{alg:EASG}
\end{algorithm}

\emph{EASG} follows a standard evolutionary algorithm scheme and is presented in Algorithm~\ref{alg:EASG}. A population of individuals evolves for a fixed number of generations. In each generation crossover and mutation operators are applied with certain probabilities, and then a population for the next generation is created by selection procedure, based on fitness function value computed for each individual.

\emph{Population.}
Each chromosome $CH_q, q=1,\ldots,population\_size$ represents some
leader's mixed strategy in the form of a vector of pure strategies
$\pi^q_i$ with their probabilities $p^q_i$:
$$CH_q = \{(\pi^q_1,p^q_1), \ldots, (\pi^q_{l_q},p^q_{l_q})\},\ \ \sum_{i=1}^{l_q}p_{l_q}^q=1,$$
where $l_q$ is the length of $CH_q$. Strategy $\pi^q_i$ is a list of leader's actions in consecutive rounds. Each chromosome in the initial population includes one randomly selected pure strategy with probability equal to $1$.

\emph{Crossover.}
A crossover operator combines two randomly chosen chromosomes by aggregating all pure strategies they contain and halving their probabilities (if a given strategy belongs to both chromosomes, the resulting probabilities are summed up). Crossover is applied to a chosen pair of chromosomes with a certain probability.

\emph{Mutation.}
In mutation operation a pair (pure strategy, round number) is uniformly selected in a chromosome. Then, starting from the selected round until the last one, a leader's action is uniformly chosen in each round (among all actions available in this round) and added to a chromosome in place of the existing action. Mutation affects each individual with a certain probability.

The role of mutation operation is to boost exploration of the leader's strategy space while crossover combines existing solutions and has more exploitation nature.

\emph{Selection.}
Chromosomes are selected to the next generation through a binary tournament with a certain selection pressure $P$, i.e. among two randomly chosen chromosomes the higher-fitted one is promoted with probability $P$ and the lower-fitted one with $1-P$. $n_e$ individuals (called \emph{elite}) from the current population with the highest fitness values are directly promoted to the next generation population (unconditionally).

\emph{Evaluation.}
The fitness function is defined as the leader's utility obtained when playing a strategy encoded by a chromosome. This utility is calculated by computing game payoffs against all possible follower's strategies and choosing the one that yields the highest value for the follower, while breaking ties in favor of the leader (SSE condition).

\subsubsection{ATSG implementation} Similarly to \emph{O2UCT}, an important advantage of \emph{EASG} formulation is its flexibility, understood as the ease of adaptation to various SG formulations. In the context of BR various types of perturbations to the optimal follower's response can be implemented in \emph{EASG} by adjusting the chromosome evaluation procedure.

Incorporation of ATSG into \emph{EASG} relies on considering a distorted version of the leader's mixed strategy when calculating the best follower's response. This distorted leader's strategy is obtained in the three following steps.
\begin{enumerate}
\item First, in order to directly apply eq.~(\ref{eq:ATSG}), a strategy encoded by a chromosome is transformed to the form of a tree.
\end{enumerate}
Formally, let's denote by $P_{pref}(\sigma)$ a sum of probabilities of all pure strategies in the chromosome with prefix $\sigma$. A probability of an edge in a game tree between nodes corresponding to $init(\sigma)$ and $\sigma$ is computed as $\frac{P_{pref}(\sigma_l)}{P_{pref}(init(\sigma_l))}$.
\begin{enumerate}
\setcounter{enumi}{1}
\item Then, all probabilities in this tree are modified according to eq.~(\ref{eq:ATSG}).
\item Finally, the tree (with changed probabilities) is transformed back to a list of pure strategies with assigned probabilities through a reversed procedure.
\end{enumerate}
Technically, each pure strategy in the resultant chromosome is created based on a unique path in the tree, from the root to a leaf node, and its probability is equal to the product of probabilities of all edges on that path. This way, the best follower's response strategy is obtained.

Next, this follower's response is used to calculate players' utilities, but this time using original, unmodified strategy from a chromosome (without distortion of probabilities).

In other words, a distorted strategy is used only in calculation of the follower's utility. The leader's strategy is assumed to be perfectly rational and therefore their utility (chromosome fitness value) is calculated with no distortion.

Similarly to \emph{O2UCT}, instead of eq.~(\ref{eq:ATSG}), the non-linear form of ATSG described in section~\ref{sec:at-modification} could be used as well. For the sake of comparison with MILP/LP methods a simplified linear ATSG formulation is considered.

\section{Experimental evaluation}%
\label{sec:evaluation}

In what follows the ATSG versions of all five considered methods will be referred to with prefix \emph{AT-}, i.e. \emph{AT-C2016}, \emph{AT-CBK2018}, \emph{AT-BC2015}, \emph{AT-O2UCT} and \emph{AT-EASG}, resp.

\subsection{Benchmark games}%
\label{sec:benchamrk}

\begin{figure}[t]
  \begin{subfigure}[t]{.48\columnwidth}
    \centering
    \resizebox{0.80\columnwidth}{!}{
    \begin{tikzpicture}[building]
\input{smallbuilding-102}
\end{tikzpicture}
    }
    \caption{An example of a warehouse layout: narrow black path denotes the main corridor, squares are storage spaces. Room numbers correspond to vertex labels in the resulting game graph presented in the right figure.}
    \label{fig:game-graph-layout}
  \end{subfigure}\quad
  \begin{subfigure}[t]{.48\columnwidth}
    \centering
    \includegraphics[width=\columnwidth]{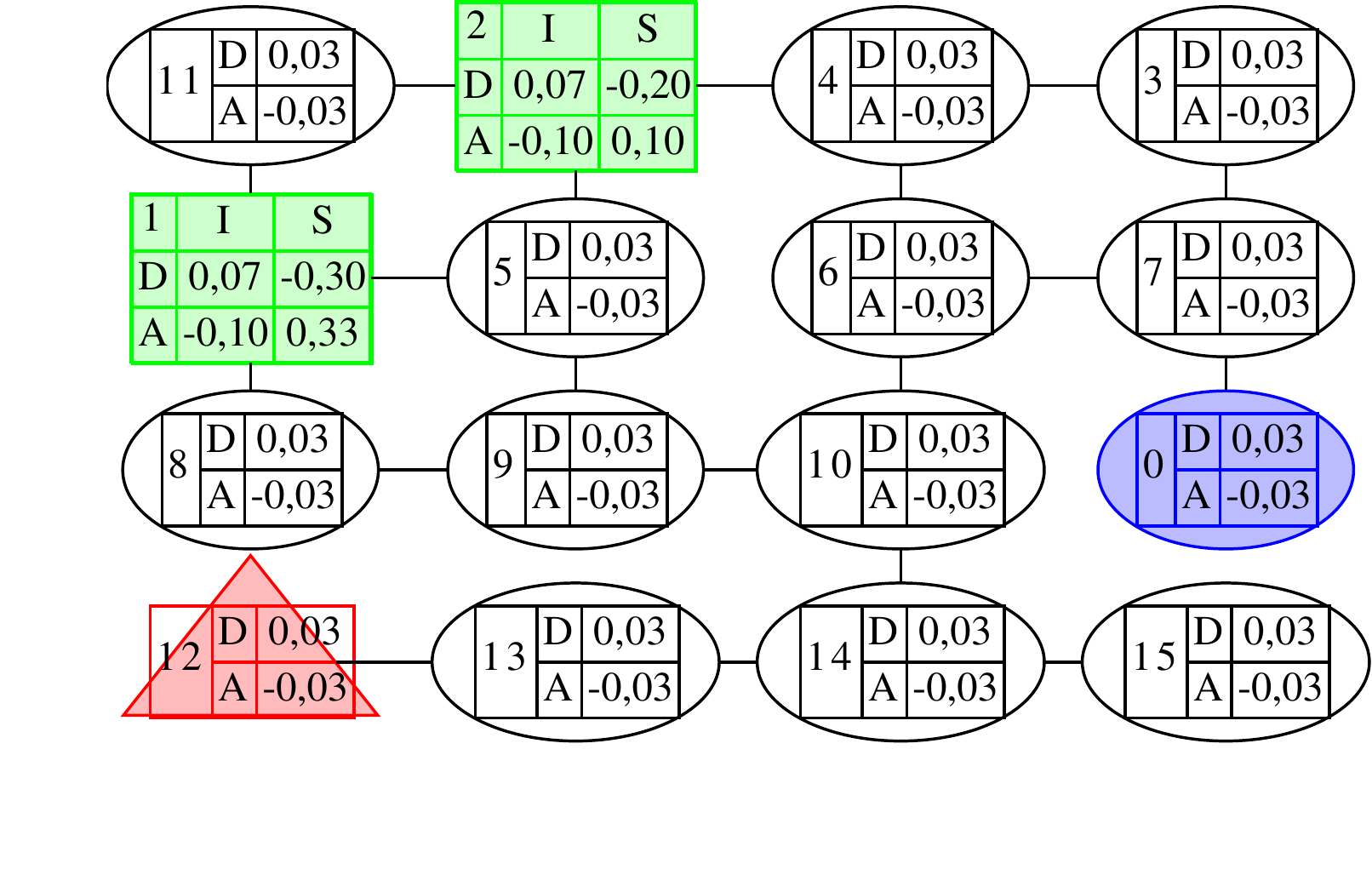}
    \caption{A corresponding game graph. Rectangular vertices are targets, a triangle vertex is attacker's starting point, a blue shaded circle vertex is defender's starting point.}
    \label{fig:game-graph-gr}
  \end{subfigure}
  \caption{An example Warehouse Game. Values in the right figure denote payoffs for the attacker and the defender, resp.\ in the case of an interception of the attacker in a given vertex. Additional utilities, in case of a successful attack, are assigned in targets (the second column). All games are defined on a $4\times 4$ grid.
  }%
  \label{fig:game-graph}
\end{figure}

Experimental evaluation was performed on a set of benchmark Warehouse Games introduced in~\cite{EJOR2019} (all game instances were downloaded from the website~\cite{sg-mini}). Each game resembles a situation of patrolling a warehouse building. The game area is modeled in the form of a graph with some vertices containing valuable resources (referred to as \emph{targets}). There are two players in a game: a defender and an attacker. Each player possesses a single unit, located in one of the vertices (warehouse spaces). In each round, each unit can either stay in a currently occupied vertex or move to an adjacent one (change the room). If the units meet in a common vertex an interception occurs and the defender receives a reward (positive utility value) while the attacker receives a penalty (negative utility). If the attacker reaches any of the target vertices (rooms) without being intercepted by the defender, he/she is rewarded and the defender is penalized. In either of the above cases the game ends. Otherwise the game is played a fixed number of rounds $T$. Once the round limit $T$ is reached, both players are assigned a neutral utility of $0$.

The benchmark set consists of $25$ games generated on $4\times 4$ grid, with general-sum utilities. An example game layout created by a warehouse generator is presented in Figure~\ref{fig:game-graph-layout} (this is an auxiliary game representation). The corresponding game graph (the actual game representation) is depicted in Figure~\ref{fig:game-graph-gr}. A detailed description of the game generator settings is presented in~\cite{EJOR2019}. In this paper games with $T=3,\ldots,7$ are considered, albeit for $T=7$ exact methods were unable to compute solutions within allotted time and memory.

\subsection{Experimental setup}%
\label{sec:setup}

For each game instance (game layout and game length) \emph{AT-O2UCT} and \emph{AT-EASG} were run $10$ times and for each other (deterministic) MILP method a single trial was performed.

Tests were run on Intel Xeon Silver 4116 @ 2.10GHz with 256GB\ RAM. Experiments with \emph{AT-O2UCT} and \emph{AT-EASG} were run in parallel, each with 8GB\ RAM assigned. Tests with \emph{AT-C2016}, \emph{AT-CBK2018}, \emph{AT-BC2015} were run sequentially with all 256GB\ RAM available in each trial. All tests were limited to $200$ hours (per single test) and were forcibly terminated if not completed within the allotted time.

Performance of both heuristic methods was analyzed in two dimensions: quality of results (an expected leader's payoff) and time efficiency. Results for all games were merged based on the number of game nodes of an extensive-form game. This grouping followed formula (\ref{eq:buckets}) presented below:
  \begin{equation}
    \label{eq:buckets}
    bucket=10^{round(\log_{10}|\mathcal{H}|)},
  \end{equation}
where
$round$ rounds a number to the nearest integer. Consequently, test games were grouped by the orders of magnitude of the respective numbers of game nodes. Such a grouping combines two aspects of game complexity: the underlying game graph structure and the game length. In the rest of the paper $B_i, i=2,\ldots,7$ will denote the \emph{$i$-th bucket of games}, i.e. the one which contains all games for which $round(\log_{10}|\mathcal{H}|)=i$. In order to streamline the notation, we will denote by $B_{\ge i}, i=2,\ldots,7$ the \emph{union of buckets} $B_i, B_{i+1},\ldots, B_7$ and by $B_{\le i}, i=2,\ldots,7$ the \emph{union of buckets} $B_2, B_3, \ldots B_i$.

The \emph{AT-BC2015} method (likewise \emph{BC2015}~\cite{bosansky2015}) is parameterless. In \emph{AT-C2016} the SI-LP variant of \emph{C2016}~\cite{cermak2016using} is considered. For \emph{AT-CBK2018}, the fast converging variant of \emph{CBK2018}~\cite{cerny2018incremental} (with $\epsilon=0.3$ and $\sigma=0.4$) is implemented.

\emph{AT-O2UCT} is parameterized by the following $3$ stopping conditions (cf. Fig. $4$ in~\cite{O2UCT}):
Either a maximum number of executions of the \emph{positive pass} (step $\dagger$ in the figure) exceeds $5\,000$, or an improvement of the leader's payoff in $500$ subsequent iterations is less than $10^{-5}$, or a number of subsequent executions of the \emph{feasibility pass} (step $\ddagger$ in the figure) without going to \emph{positive pass} (step $\dagger$) exceeds $10\,000$ (\emph{infeasible strategy}).

In \emph{AT-EASG} the following values for the steering parameters are selected: population size - $30$, mutation probability - $0.5$, crossover probability - $0.8$, selection pressure $P=0.9$, the number of elitist chromosomes - $n_e=2$. The algorithm is run either for $1\,000$ generations or until no improvement of the leader's strategy is observed in $20$ subsequent generations (whichever occurs first).

Parameters for the last two methods were selected based on a limited number of preliminary simulations.

\subsection{Payoffs}%
\label{sec:payoffs}

\begin{figure}
  \centering
    \includegraphics[width=.80\columnwidth]{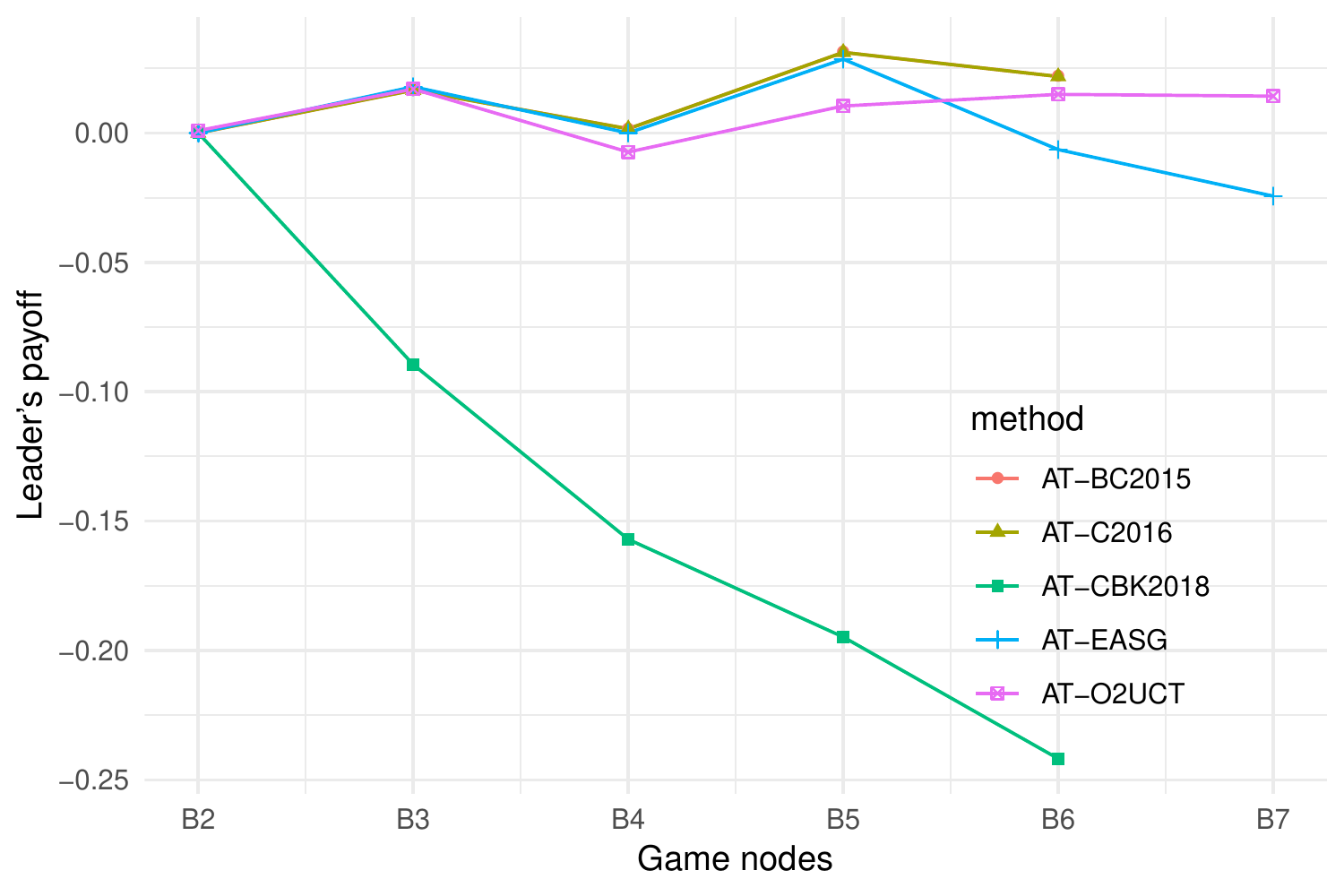}
  \caption{The average expected leader's utility.}%
  \label{fig:payoff}
\end{figure}

The average expected utilities of the leader obtained by each method are presented in Fig.~\ref{fig:payoff}. Since both \emph{AT-C2016} and \emph{AT-BC2015} are exact methods their results are clearly the highest and the respective plots overlap. Both non-MILP heuristic methods perform slightly worse, although
\emph{AT-EASG} is a close runner-up for games from $B_{\le 5}$, in which range it outperforms \emph{AT-O2UCT}.

For the largest games, from $B_7$, the best-performing method is \emph{AT-O2UCT}, which excels \emph{AT-EASG} (the only remaining competitor) by a clear margin. None of the two exact MILP methods were capable of solving games of this size (belonging to $B_7$) and the approximate MILP approach (\emph{AT-CBK2018}) solved $16$ game instances and failed in solving the remaining $9$. Consequently, for the sake of fair comparison, the results of \emph{AT-CBK2018} are not presented for the largest games.

Generally speaking \emph{AT-CBK2018} yields the weakest outcomes across the entire range of game sizes and its performance deteriorates along with increasing game complexity.

\subsection{Time scalability}%
\label{sec:times}

\begin{figure}
  \centering
    \includegraphics[width=.80\columnwidth]{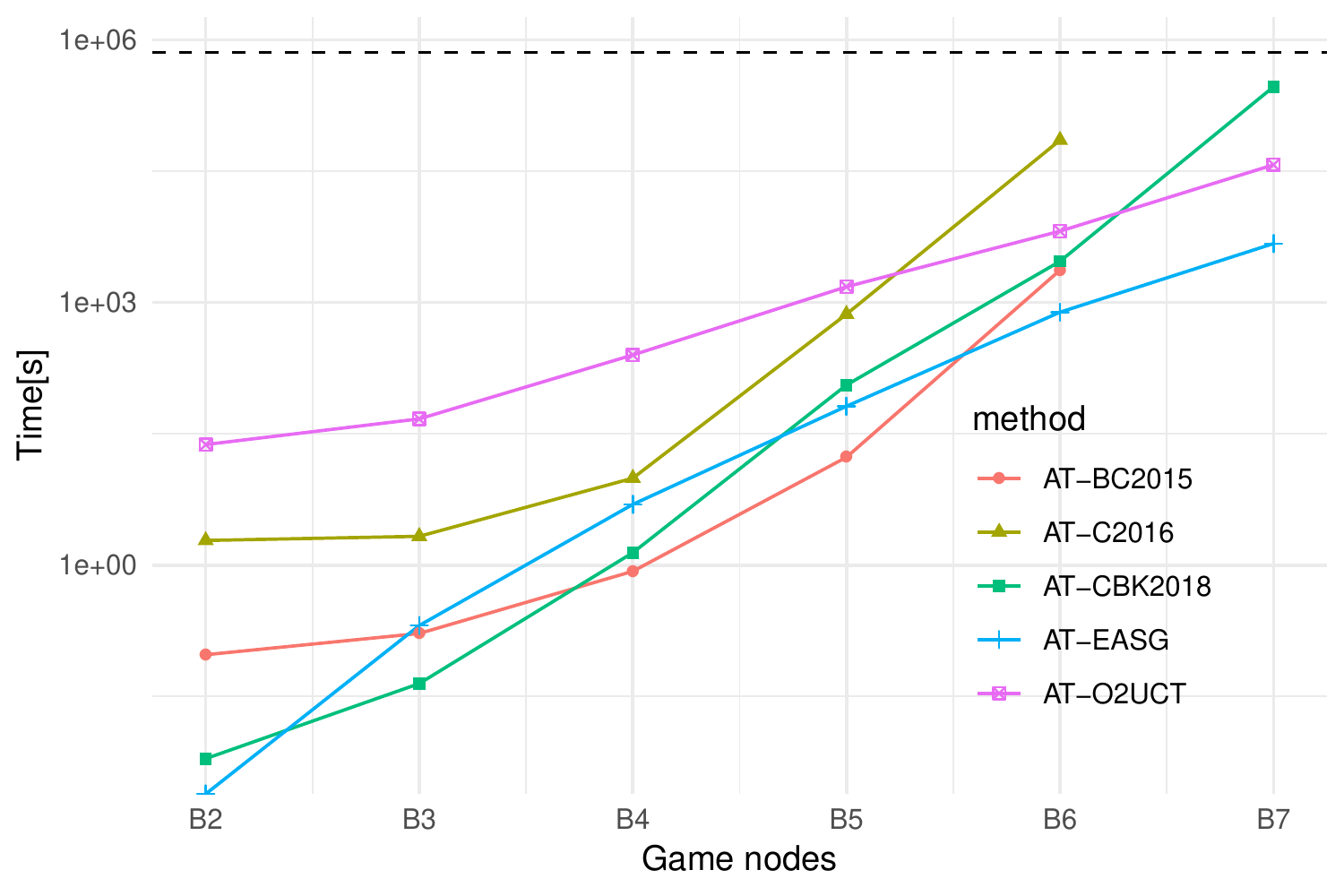}
  \caption{The average time requirements.}%
  \label{fig:time}
\end{figure}

Fig.~\ref{fig:time} presents time scalability of the methods. While all of them scale exponentially, the running times of both non-MILP approaches grow at slower paces. For games from $B_{\ge 6}$ (\emph{AT-EASG}) and from $B_{\ge 7}$ (\emph{AT-O2UCT}), resp. they already excel both exact MILP methods.

On the other hand, it shouldn't be forgotten that the main asset of \emph{AT-C2016} and \emph{AT-BC2015} is convergence to the optimal solution and therefore a comparison of their running times with heuristic approaches needs to be considered with care.

Nevertheless, it seems reasonable to conclude that beyond certain level of game complexity
the exact methods become infeasible and, in such scenarios, both heuristic approaches present a viable alternative.

The third MILP method is a state-of-the-art algorithm for approximate solving extensive-form games. Following a recommendation from~\cite{cerny2018incremental} \emph{AT-CBK2018} was parameterized in a way which assures fast convergence ($\epsilon=0.3, \sigma=0.4$) so as to make a fair time comparison with the remaining two heuristic methods.
It can be concluded from Fig.~\ref{fig:time} that for the set of most complex games \emph{AT-EASG} and \emph{AT-O2UCT} are faster than \emph{AT-CBK2018}, at the same time providing much better leader's payoffs (cf. Fig.~\ref{fig:payoff}).

Please note that since \emph{AT-CBK2018} solved only $16$ games from $B_7$, the solution times for the remaining game instances were capped at the limit of $200$h, which was in favor of \emph{AT-CBK2018} (in comparison with \emph{AT-O2UCT} and \emph{AT-EASG}) since both heuristic methods solved all $25$ largest games within the allotted time (hence, in their case the actual times are reported).

\subsection{Stability of \emph{AT-O2UCT} and \emph{AT-EASG}}%
\label{sec:EASG}

\begin{figure}[t]
  \centering
    \includegraphics[width=.80\columnwidth]{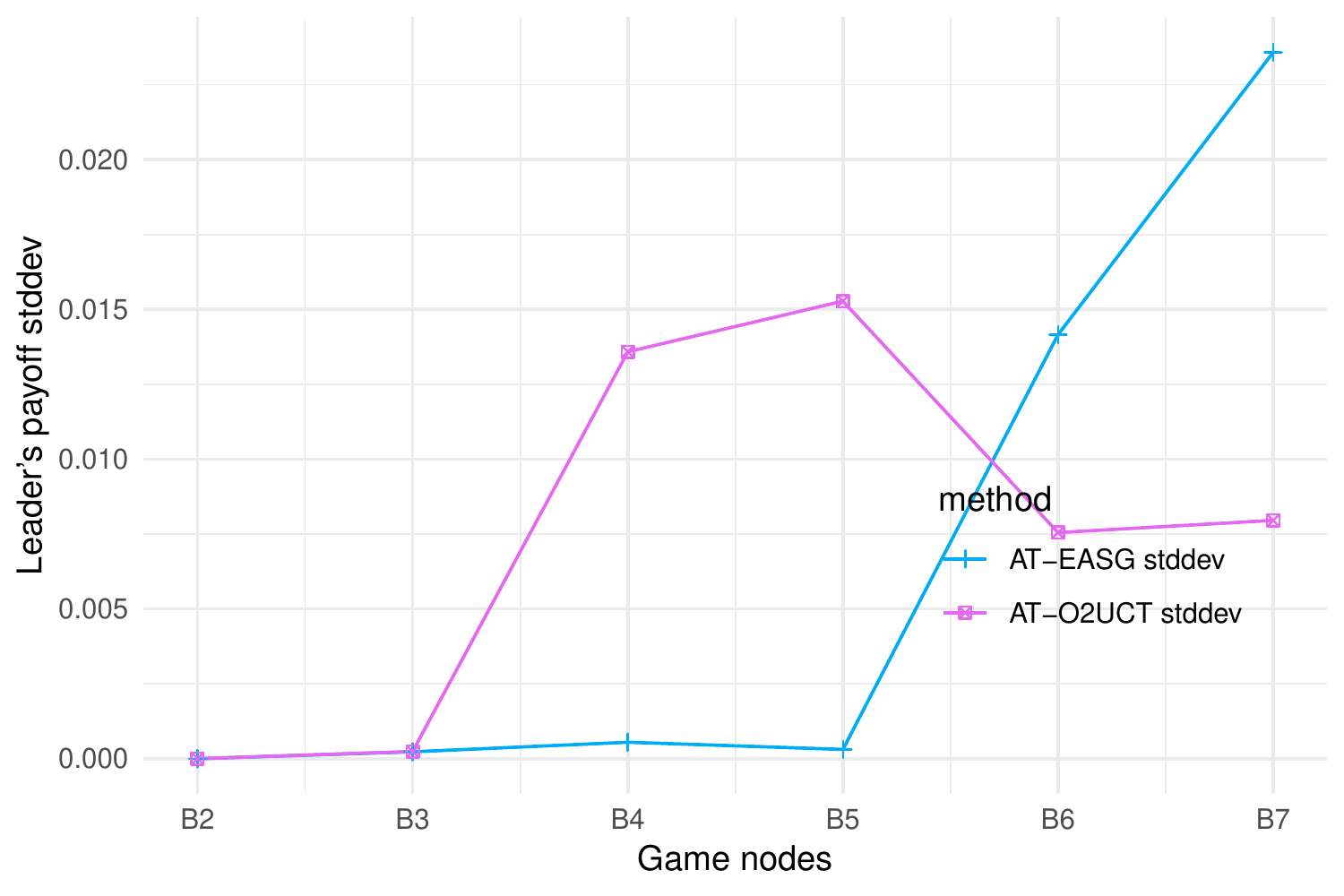}
  \caption{The average standard deviation of the expected leader's utility for \emph{AT-O2UCT} and \emph{AT-EASG} methods.}%
  \label{fig:payoff-sd}
\end{figure}

\begin{figure}[h!]
  \centering
    \includegraphics[width=.80\columnwidth]{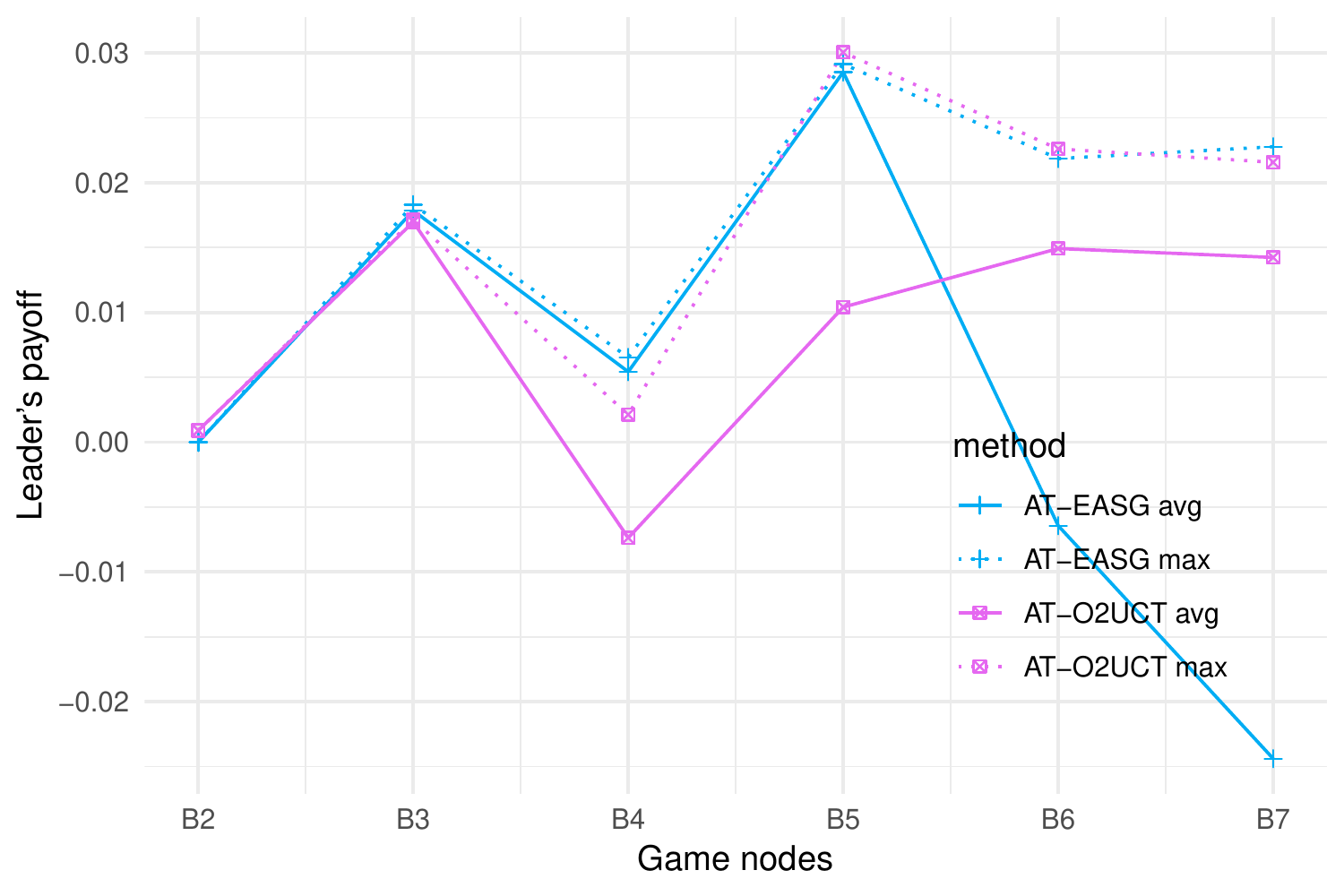}
  \caption{The average and maximum of the expected leader's utility for \emph{AT-O2UCT} and \emph{AT-EASG} methods.}%
  \label{fig:payoff-max}
\end{figure}

While both non-MILP methods proved efficient in both time scalability and returned payoffs, one of the interesting observations from Fig.~\ref{fig:payoff} is deterioration of the average payoffs obtained by \emph{AT-EASG} for the largest games (from $B_{\ge 6}$). Apparently, along with further increasing of games' complexity the variance of \emph{AT-EASG} results also increases. In particular, for two game instances from the considered benchmark set~\cite{sg-mini} (\emph{smallbuilding-89-6} and \emph{smallbuilding-89-7}) standard deviations of results equaled nearly $0.26$ which had a visible impact on the average payoffs. Figure~\ref{fig:payoff-sd} compares the average standard deviations (stddev.) for \emph{AT-EASG} and \emph{AT-O2UCT}. Clearly, for games from $B_{\ge 6}$ \emph{AT-EASG} stddev. increases, while stddev. of \emph{AT-O2UCT} remains approximately on the same level. In terms of stability, \emph{AT-O2UCT} appears to have a clear advantage over \emph{AT-EASG}.

On the other hand, despite high variance, \emph{AT-EASG} is still able to obtain very good solutions, with the best ones practically equal to those of \emph{AT-O2UCT}. Figure~\ref{fig:payoff-max} shows both the average and maximum payoffs of both methods. Dotted curves (which represent the maximum payoff averaged across all benchmark games from a given bucket) are very close to each other and the differences between the best solutions found by \emph{AT-O2UCT} and \emph{AT-EASG} are below $10^{-2}$ for each bucket.

\section{Conclusions}%
\label{sec:conclusions}

This work considers the SG formulation in which the follower is not perfectly rational. Such a setting is motivated by the real SG scenarios in which humans, when performing the role of the follower, are prone to certain inefficiencies in perception and assessment of the leader's strategy. A particular implementation of the follower's bounded rationality considered in this paper refers to Anchoring Theory~\cite{tversky1974judgment}. AT assumes the existence of a certain distortion (towards the uniform distribution of probabilities of possible actions) of the follower's perception of the leader's mixed strategy. The leader being aware of that distortion can exploit this weakness in their strategy formulation.

The paper proposes an efficient MILP-suitable formulation of AT in the context of sequential extensive-form SGs. This formulation (ATSG) is implemented in three state-of-the-art MILP methods -- two exact ones: \emph{BC2015}~\cite{bosansky2015} and \emph{C2016}~\cite{cermak2016using} and one approximate: \emph{CBK2018}~\cite{cerny2018incremental}, as well as in two heuristic non-MILP approaches: \emph{O2UCT}~\cite{KarwowskiMandziuk2019,O2UCT} and \emph{EASG}~\cite{EASG}.

Experimental results on a set of $25$ games show that non-MILP methods provide optimal or close-to-optimal leader's payoffs while being visibly faster than exact MILP approaches. At the same time, they clearly outperform time-optimized approximate MILP method in both payoffs quality and time efficiency.

An additional asset of non-MILP solutions in the context of BR is their flexibility, stemming from virtually no restrictions imposed on the form in which BR is represented. Unlike MILP methods which require a linear form of BR related constraints, non-MILP solutions allow the implementation of more complex, non-linear BR formulations.

Our current focus is on verification of the efficacy of proposed AT formulation in experiments that involve human players in the role of the follower. To this end
for a selection of Warehouse Games the following leader's strategies will be precomputed: (a) SSE strategy (rational, not distorted), (b) ATSG SSE strategy and (c) a strategy stemming from eq.~(\ref{eq:ATSG-nl}) approximated by \emph{AT-EASG} or \emph{AT-O2UCT}, respectively. In each game, a human participant will play the follower's role against (randomly assigned) one of the above strategies. A comparison of the average leader's payoffs in case (a) vs (b) and (c) will indicate whether potential distortions of SSE can, in fact, provide any advantage for the leader, stemming from the fact that the role of the follower is indeed played by a human.

\section*{Acknowledgments}
The work was supported by the National Science Centre grant number\linebreak[4]
2017/25/B/ST6/02061.

\bibliographystyle{plain}
\bibliography{modanchoring}

\end{document}